\documentclass[preprint,showpacs]{revtex4} 

\usepackage{graphicx}
\usepackage{amssymb}

\begin{document}

\title{Decoupling of diffusion from structural relaxation and
spatial heterogeneity in a supercooled simple liquid}

\author{Mikhail Dzugutov, Sergei I. Simdyankin, Fredrik H. M. Zetterling}

\affiliation{Department of Numerical Analysis, 
Royal Institute of Technology, SE-100 44 Stockholm, Sweden}

\date{\today}

\begin{abstract} 
We report a molecular dynamics simulation of a supercooled simple
monatomic glass-forming liquid. It is found that the onset of the
supercooled regime results in formation of distinct domains of slow
diffusion which are confined to the long-lived icosahedrally
structured clusters associated with deeper minima in the energy
landscape. As these domains, possessing a low-dimensional geometry,
grow with cooling and percolate below $T_c$, the critical temperature
of the mode coupling theory, a sharp slowing down of the structural
relaxation relative to diffusion is observed. It is concluded that
this latter anomaly cannot be accounted for by the spatial variation
in atomic mobility; instead, we explain it as a direct result of the
configuration-space constraints imposed by the transient structural
correlations. We also conjecture that the observed tendency for
low-dimensional clustering may be regarded as a possible mechanism of
fragility.
\end{abstract}

\maketitle

Fragile liquids\cite{Angell91}, having been cooled below a
characteristic temperature, $T_A$, which is typically close to the
melting point, undergo a transition to the supercooled dynamics regime
with super-Arrhenius slowing down and stretched exponential
relaxation. Mode-coupling theory \cite{Gotze92} appears only to be
successful in interpreting early stages of supercooled
dynamics. Further cooling results in a fundamental transformation of
the liquid state that has not yet been comprehended in terms of
theoretical models \cite{Ediger96}. This transformation is manifested
by three principal phenomena observed in the vicinity of the glass
transition point $T_g$: (i) the liquid undergoes a structural
transformation shifting to the area of its energy landscape with
deeper minima \cite{Sastry98} (ii) a long-range slowly relaxing
spatial heterogeneity arises\cite{Sillescu99} that is observed as
formation of structurally\cite{Weeks00} and
dynamically\cite{Glotzer98} distinct long-lived domains (iii) a new
type of liquid dynamics develops where the structural relaxation
becomes retarded relative to the translational diffusion, thus
breaking the Stokes-Einstein relation \cite{Blackburn96}. It
appears sensible to ask whether these three observations represent
different aspects of the same phenomenon, and, if so, what is its
primary mechanism.

Here, we address these questions in a molecular dynamics simulation
examining the structural and dynamical aspects of a simple monatomic
liquid in a strongly supercooled equilibrium state. The model
comprises 16000 particles interacting via a pair potential
\cite{Dzugutov92} designed to favour icosahedral order in the first
coordination shell. In this way, the liquid imitates the structure of
simple metallic glass-formers \cite{Hafner}. Sufficiently long
relaxation under supercooling transforms it into a dodecagonal
quasicrystal \cite{Dzugutov93}; but, due to its exceeding complexity,
this transformation can be delayed, keeping the liquid in a metastable
supercooled state on a time-scale that allows us to explore its
essential dynamical properties \cite{Dzugutov94}. The potential and
all the quantities computed here are expressed in terms of the
Lennard-Jones reduced units. We cool the liquid towards the glass
transition at a constant density $\rho=0.85$ in a step-wise manner,
equilibrating it at each temperature point.

\begin{figure} 
\includegraphics[width=6.5cm]{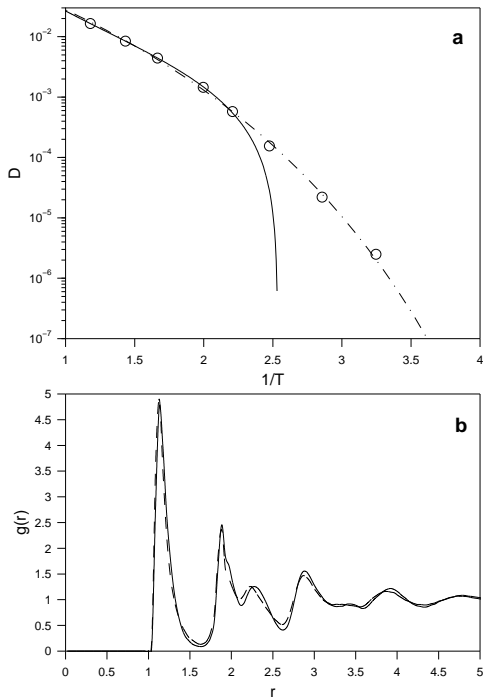}
\caption{(a) Open circles: temperature variation of the diffusion
coefficient $D$.  Deviation from the Arrhenius behaviour is observed
around $T=0.8$. Solid line is a fit of the equation:
$D=A(T-T_c)^{\gamma}$ which yields the value of the mode-coupling
theory critical temperature $T_c=0.39$. Chain-dashed line is 
the fit of Vogel-Fultcher-Tamman law: $D= D_0\exp (B T_0/(T-T_0))$,
with $T_0=0.17$ and $B=9.14$. Parameter $B$ is an indicator of the
liquid's fragility \cite{Angell91}. (b) Radial distribution functions
for two IS configurations. Solid line: $T=0.3$; dashed line: $T=1.0$.}
\end{figure}

The temperature variation of the self-diffusion coefficient is shown
in Fig. 1a. A clear transition to the supercooled dynamics regime
marked by the onset of super-Arrhenius behaviour occurs at
$T_A=0.8$. This is accompanied by a pronounced stretched-exponential
relaxation \cite{Dzugutov94}. We analyse the energy landscape
transformation under cooling by performing the steepest descent energy
minimization of the instantaneous liquid configurations \cite{jncs}
producing the so-called inherent structure (IS) configurations
\cite{Stillinger84}. In spite of the dramatic drop of diffusivity
between $T=1$ and $T=0.3$, the respective IS radial distribution
functions shown in Fig.1b indicate only a marginal change of the local
order. Nevertheless, the IS energy (Fig. 2a), remaining almost
constant at higher temperatures, decreases as the liquid is cooled
below $T_A$. A similar effect was observed in a supercooled
two-component Lennard-Jones liquid \cite{Sastry98}.

The principal issue that we address in this study concerns the effect
that the energy landscape transformation as indicated by the reduction
of IS energy has on the liquid dynamics. In this context, we
investigate two dynamical anomalies mentioned above: spatial variation
in atomic mobility and breaking the Stokes-Einstein relation. It is
intuitively clear that the former must be intimately connected with
the structural heterogeneity. It was suggested \cite{Kirkpatrick89}
that a liquid approaching $T_g$ develops extensive domains of distinct
structure based on the energy-favoured local order incompatible with
periodicity, which is apparently icosahedral in the present case. A
detailed investigation of the evolution of icosahedral order in this
system under cooling can be found elsewhere \cite{jncs, Dzugutov92};
its results essential for the present analysis are shown in
Fig. 2. The number of icosahedrally coordinated atoms, Fig. 2b,
remains almost temperature-independent for $T>T_A$, and grows rapidly
with cooling below $T_A$. This clearly correlates with the behaviour
of the IS energy in Fig. 2a, indicating that the lower-energy minima
in the energy landscape that the liquid occupies in the supercooled
state are associated with icosahedral order.

Fig. 2c shows the temperature variation of the maximum size of
continuous aggregations of connected icosahedra (it is assumed that
two icosahedra sharing at least 3 atoms are connected). These
low-energy configurations possess a higher mechanical stability than
other configurations arising in the liquid. The pattern in Fig. 2c is
consistent with those shown in Figs. 2a and 2b. The cluster size
remains temperature-independent above $T_A$ and grows rapidly with
cooling below $T_A$, diverging around $T_c$ where connected icosahedra
form a percolating network.  We also note that the onset of
percolation reduces the temperature variations of both the number of
icosahedra and the IS energy.

The growth of a continuous bulk aggregation of icosahedrally
coordinated atoms in a flat 3D space is limited by the rapidly
increasing strain energy caused by geometric frustration. Here, the
unlimited size of the icosahedral clusters is a result of
low-dimensional growth. An example demonstrating this tendency - a
cluster of 716 atoms detected at $T=0.45$ - is shown in Fig. 2d. The
icosahedral clusters with low-dimensional geometry were observed in
the global energy-minima analysis of this system\cite{dw}. This
analysis also demonstrated that the effect of low-dimensional
icosahedral aggregation is destroyed by a variation of the pair
potential \cite{wales01}. It is worth noting that although icosahedral
coordination was found to be ubiquitous for the energy-minima clusters
of the Lennard-Jones system \cite{wales97}, in agreement with an
earlier study \cite{Jonsson88}, these clusters demonstrate a
distinctly bulk pattern of aggregation.

\begin{figure} 
\includegraphics[width=8.2cm]{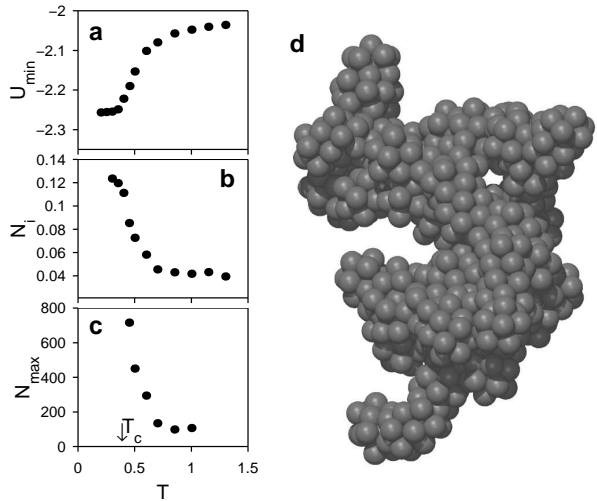}
\caption{(a) Temperature variation of the IS energy, per atom. (b)
Temperature variation of the relative number of atoms with icosahedral
coordination (we assume two atoms separated by a distance of less than
1.5 to be neighbours) (c) Temperature variation of the maximum size of
cluster of interconnected icosahedra. The error bars in plots (a)-(c),
as estimated from 3 independent runs, are of order of the dots size
(d) The largest icosahedral cluster detected at $T=0.45$ comprising
716 atoms. Note that its size exceeds the range of structural
correlations as estimated from Fig.~1b.}
\end{figure}

\begin{figure} 
\includegraphics[width=7.9cm]{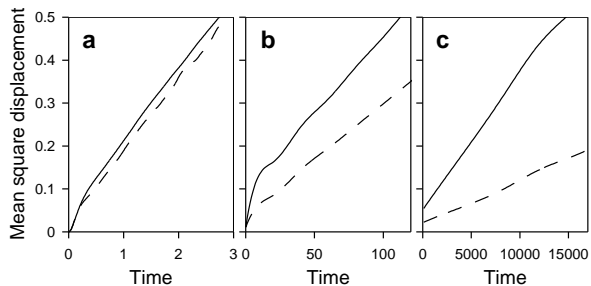}
\caption{The impact of the structural
heterogeneity on diffusion. Dashed lines and solid lines denote,
respectively, mean-square displacement calculated for the atoms that
were icosahedrally coordinated at the initial moment of time, and for
those with non-icosahedral environment.  (a) $T=1.0$, (b) $T = 0.45$,
and (c) $T=0.3$.}
\end{figure}

\begin{figure} 
\includegraphics[width=5. cm]{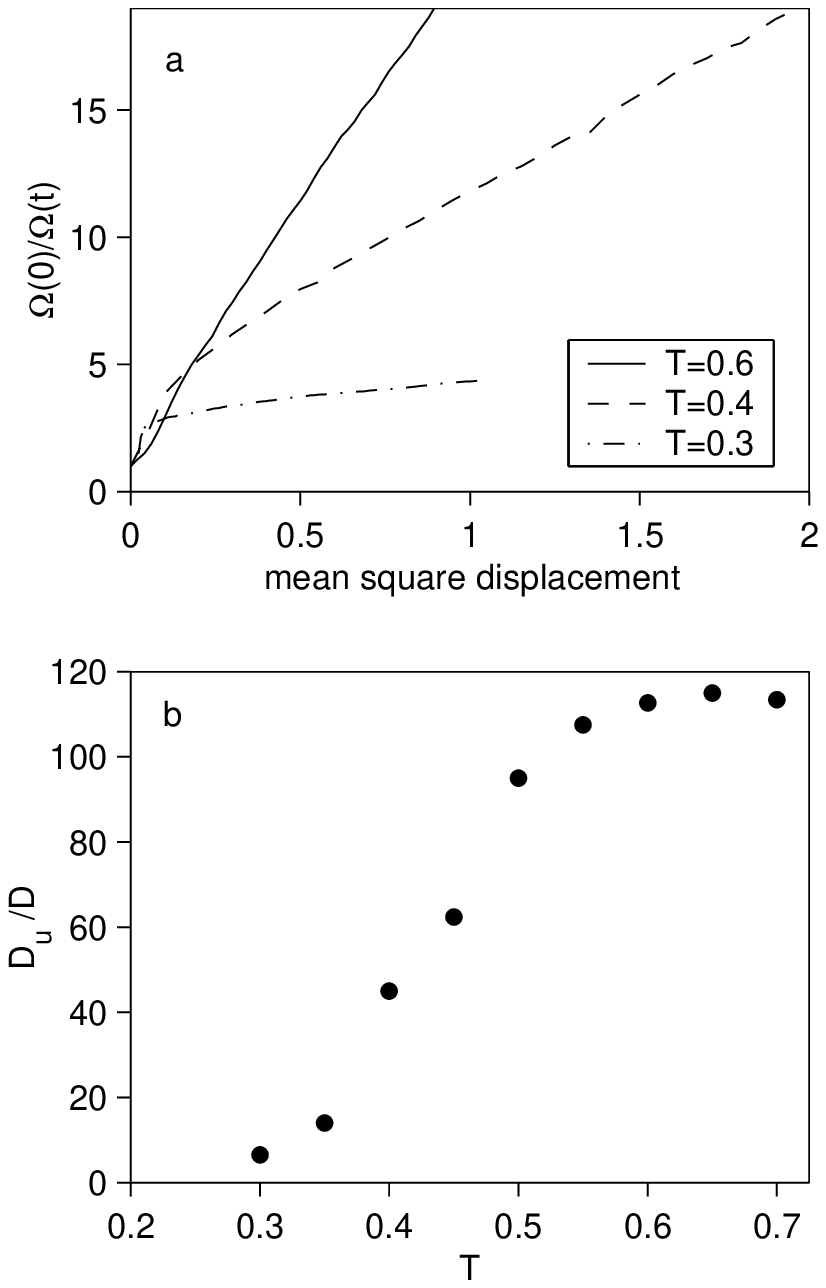}
\caption{(a) $\Omega_u^{-1}$ as a function of mean-square
displacement. $\Omega_u$ is defined by Eq. 1 (b) Temperature variation
of $D_u/D$. The latter is obtained as the slope in asymptotically
linear variation of $\Omega^{-1}$ as a function of mean-square
displacement. Note that at $T=0.3$ the system is apparently
non-equilibrated}
\end{figure}

This structural heterogeneity is accompanied by the dynamical
heterogeneity. In Fig. 3, the mean-square displacement (MSD) for the
atoms initially situated inside the icosahedral domains is compared
with that for the initially external atoms with non-icosahedral
environments. As the domain size grows under cooling, the latter group
of atoms becomes increasingly more mobile than the former one. These
results are consistent with earlier observations associating low
mobility domains in supercooled liquids with lower energy 
{\cite{donati} and higher degree of local ordering {\cite{harrow}.

Next, we investigate the relation between diffusion and structural
relaxation. The rate of ergodicity-restoring structural relaxation can
be assessed by comparing the time-average of a local variable with its
ensemble-average \cite{Ma,Mountain89}. If $u_i$ is the potential
energy of atom $i$, the respective measure for a system of $N$ atoms
is:
\begin{equation}
\Omega_u (t) = \frac{1}{N} \sum_{i =1}^N \left [ \frac{1}{t} \int_0^t
u_i(s) ds -\langle u_i \rangle \right ]^2
\end{equation}
where $\langle \ \rangle$ denotes the ensemble average. This measure
decays with time \cite{Mountain89} as:
\begin{equation}
\Omega_u (t) = \Omega_0 / D_u t
\end{equation}
where $D_u$ is a constant, and $\tau = D_u^{-1}$ can be regarded as
characteristic relaxation time. Fig. 4(b) shows the temperature
variation of $ D_u/D$ (the latter was estimated from the slope of
asymptotically linear variation of $\Omega_u^{-1}$ as a function of
MSD, Fig. 4(a)) It remains constant in the stable liquid domain and
decreases rapidly as the liquid is cooled below $T_A$. The steepest
variation of $D_u/ D$ occurs around $T_c$; a tendency for saturation
observed in the $T_g$ area possibly indicates that the system remains
non-equilibrated. This pattern is thus perfectly consistent with the
temperature variations of the IS energy and the domain structure shown
in Fig. 2.

Two distinct aspects of dynamical heterogeneity are commonly discussed
in relation with breaking the Stokes-Einstein relation
\cite{cicerone}. One model \cite{hodg} explains the latter anomaly as
a direct result of the existence of spatial domains with distinctly
different rate of atomic mobility. It conjectures that the
translational diffusion is mostly confined to the ``fast'' domains,
while the structural relaxation (viscosity) is controlled by the
reduced mobility in the ``slow'' domains; at the same time, the
dynamics in each separate domain is assumed to be adequately described
by the Stokes-Einstein relation \cite{Blackburn96}. We note that
although the spatial variation in atomic mobility is found in the
present study, Fig. 3, the scale of this effect is by more than an
order of magnitude smaller than that of the relaxation-diffusion
decoupling in Fig. 4. This observation clearly demonstrates that even
in the domains with lowest atomic mobility diffusion becomes strongly
enhanced with respect to structural relaxation, which is obviously
inconsistent with the above model.

Another aspect of dynamical heterogeneity that has been conjectured to
explain the Stokes-Einstein breaking is dynamical cooperativity
\cite{fujara}. We present here some arguments in favour of this
conjecture. First, we have to understand why diffusivity and the rate
of structural relaxation remain universally connected in the stable
liquid state. Let us consider a coarse-grained configuration-space
trajectory of a liquid system with with a certain finite value of MSD
per an elementary step. We assume that the relaxation process of a
stable liquid represents a random walk in the configuration space
constrained by the ensemble-averaged structural correlations. This
assumption implies that as a result of an elementary step the system
can, with equal probability, be found in any available configuration
space point within the indicated MSD range from its initial position
(availability implies that the configuration is allowed by the
equilibrium ensemble-averaged structural correlations)\cite{nat}. It
is clear that the ratio of the average number of available
configurations that the system can access in a relaxation step with a
fixed MSD to the total number of available configurations remains
constant within the domain of stable liquid state where the above
assumption is assumed to be valid. On the other hand, this ratio can
be regarded as a measure of the relaxation rate if time is expressed
in terms of MSD.

Next, we assume that the relaxation process is additionally constrained
by time-limited correlations complementary to the ensemble-average
equilibrium structural correlations. These additional correlations are
characteristic of the supercooled liquid state; they can be observed
both as positional correlations in the form of a transient domain
structure discussed above and as a long-range dynamical cooperativity
\cite{Palmer84}. In this case, the system in its elementary relaxation
step cannot access all the available configuration points within the
respective MSD range. As a result, diffusion becomes enhanced with
respect to the structural relaxation. The magnitude of this
effect is apparently controlled by the scale of the described
time-limited complementary correlations. Indeed, the dramatic decrease
of $D_u/D$ in Fig. 4. clearly correlates with the divergence of the
correlation length associated with cluster percolation at $T=T_c$
(Fig. 2c).

In conclusion, the results presented here connect the profound change
in the liquid's behaviour at $1.2T_g \approx T_c$ to the
transformation of its residence area on the energy landscape. The
coherently structured domains associated with low-energy minima which
are occupied in the supercooled regime represent free-energy barriers
dividing the phase-space into components \cite{Palmer82}. In this
landscape, the ergodicity restoring relaxation is facilitated by
strongly correlated atomic motions \cite{Parisi} which, as we have
shown, are inefficient in exploring the configuration space. We note
that the percolation of icosahedral order observed here resembles the
picture of rigidity percolation in bonded glass-formers
\cite{Phillips85}. An interesting question that can be addressed in a
separate study is whether this percolation occurs at the same
temperature as the percolation transition for the higher-mobility
domains which too was found around $T_c$ \cite{donati}. Another remark
concerns a possible connection between the domain geometry and
fragility. The latter is related to the steepness of the slowing down
of structural relaxation relative to diffusion shown in Fig.4 which is
concluded to be controlled by the domain growth rate. The
low-dimensional domain geometry avoids the inherent geometric
frustration that limits the bulk growth of domain structure and, in
this way, facilitates rapid increase of the domain size. Therefore, it
could conceivably be regarded as a generic feature of the fragile
glass-formers. Indeed, structurally and dynamically distinct
low-dimensional domains have been found in a supercooled two-component
Lennard-Jones liquid\cite{Jonsson88,Glotzer98}.

We thank C.A. Angell, S.R. Elliott and D.J. Wales for valuable
comments. We used VMD software \cite{VMD}. This study was supported by
Swedish Research Council (VR).


\begin{thebibliography}{199}
\bibitem{Angell91}
C. A. Angell, 
{\it J. Non-Cryst. Solids} {\bf 131-133,} 13 
(1991). 

\bibitem{Gotze92}
W. G\"otze,
L. Sj\"ogren,
{ Rep. Prog. Phys.} {\bf 55,} 241 
(1992).

\bibitem{Ediger96}
M. D. Ediger, 
C. A. Angell, 
S. R. Nagel, 
{ J. Phys. Chem.} {\bf 100,} 13200 
(1996).

\bibitem{Sastry98}
S. Sastry, 
P. G. Debenedetti, 
F. H. Stillinger, 
{ Nature} {\bf 393,} 554 
(1998). 


\bibitem{Sillescu99}
H. Sillescu,
{ J. Non-Cryst. Solids} {\bf 243,} 81 
(1999).

\bibitem{Weeks00} 
E. Weeks, 
J. C. Crocker, 
A. Levitt, 
A. Schofield, 
D. A. Weitz, 
{ Science} {\bf 287}, 627 
(2000).

\bibitem{Glotzer98} 
W. Kob { et al.}, 
{ Phys. Rev. Lett.} {\bf 80}, 2338 
(1998)

\bibitem{Blackburn96} 
F. R. Blackburn, 
C.-Y. Wang, 
M. D. Ediger, 
{ J. Phys. Chem.} {\bf 100,} 18249 
(1996);
I. Chang, 
H. Sillescu,
{ J. Phys. Chem. B} {\bf 101,} 8794 
(1997);
A. Mukherjee, S. Bhattacharyya and B. Bagchi, 
{J. Chem. Phys.} {\bf
116}, 4577 (2002)

\bibitem{Dzugutov92}
M. Dzugutov, 
{ Phys. Rev. A} {\bf 46,} R2984 
(1992). 

\bibitem{Hafner} 
J. Hafner, 
{ From Hamiltonians to phase diagrams} 
(Springer-Verlag, Berlin, 1988).

\bibitem{Dzugutov93}
M. Dzugutov, 
{ Phys. Rev. Lett.} {\bf 70,} 2924 
(1993).

\bibitem{Dzugutov94}
M. Dzugutov, 
{ Europhys. Lett.} {\bf 26,} 533 
(1994).

\bibitem{jncs}
F. Zetterling, M. Dzugutov, and S. Simdyankin, 
{ Journ. Non-Cryst. Solids} {\bf 293-295,} 39 
(2001).

\bibitem{Stillinger84}
F. H. Stillinger, 
T. A. Weber, 
{ Science} {\bf 225,} 983 
(1984).

\bibitem{Kirkpatrick89}
T. R, Kirkpatrick, 
D. Thirumalai, 
P. G. Wolynes, 
{ Phys. Rev. A} {\bf 40,} 1045 
(1989);
E. W. Fischer, 
E. Donth, 
W. Steffen, 
{ Phys. Rev. Lett.} {\bf 68,} 2344 
(1992);
D. Kivelson, 
S. A. Kivelson, 
X. Zhao, 
Z. Nussinov, 
G. Tarjus, 
{ Physica A} {\bf 219,} 27 
(1995).

\bibitem{dw} 
J.P.K. Doye, D.J. Wales, and S.I. Simdyankin 
Faraday Discuss., {\bf 118}, 159
(2001). 

\bibitem{wales01} 
J.P.K. Doye and D.J. Wales, Phys. Rev. Lett., 86, 5719-5722 (2001)

\bibitem{wales97} 
D.J. Wales and J.P.K. Doye, J. Phys. Chem. A, 101, 5111 (1997)

\bibitem{Jonsson88} 
H. Jonsson, 
H. C. Andersen, 
{ Phys. Rev. Lett.} {\bf 60}, 2295 
(1988).

\bibitem{donati} C. Donati, S. Glotzer, P. Poole, W. Kob and
S. Plimpton, Phys. Rev. E {\bf 60}, 3107 (1999) 

\bibitem{harrow}M. Hurley and P. Harrowell, Phys. Rev. E {\bf 52},
1694 (1995)

\bibitem{Ma} 
S. K. Ma, 
{ Statistical mechanics}, (World Scientific, Singapore, 1990).

\bibitem{Mountain89}
R. D. Mountain, 
D. Thirumalai, 
{ J. Phys. Chem.} {\bf 93,} 6975 
(1989). 

\bibitem{cicerone} 
M. Cicerone, F. Blackburn, and M. Ediger,
J. Chem. Phys., {\bf 102}, 471 (1995)


\bibitem{hodg} 
F. Stillinger and J.Hodgdon, Phys.Rev. E {\bf 50}, 2064 (1994);
M. Cicerone and M. Ediger, J. Chem. Phys. {\bf 104}, 7210 (1996)


\bibitem{fujara} F. Fujara, B. Geil, H. Sillescu and G. Fleischer,
Z. Phys. B {\bf 88} 195 (1992)

\bibitem{nat}
M. Dzugutov,
Phys. Rev. E, {\bf 65}, 032501 (2002);
M. Dzugutov, 
{ Nature} {\bf 381}, 137 
(1996);

\bibitem{Palmer84}
R. G. Palmer, 
D. L. Stein, 
P. W. Anderson, 
{ Phys. Rev. Lett.} {\bf 53,} 958 
(1984);
C. Bennemann, 
C. Donati, 
J. Baschnagel, 
S. Glotzer, 
{ Nature} {\bf 399,} 246 
(1999).

\bibitem{Palmer82} 
R. G. Palmer, 
{ Adv. Phys.} {\bf 31,} 669 
(1982).

\bibitem{Parisi} A. Barrat, S. Franz, G. Parisi, { J.Phys. A:
Math. Gen.} {\bf 30,} 5593, (1997); K. Fuchisaki, K. Kawasaki, {
J. Phys. Soc. Japan} {\bf 67,} 2158 (1998); T. Schroeder, S. Sastry,
J. Dyre and S. Glotzer, {J. Chem. Phys} {\bf 112}, 9834 (2000)


\bibitem{Phillips85} 
J. C. Phillips, 
M. F. Thorpe, 
{ Sol. State Commun.} {\bf 53}, 699 
(1985); C. A. Angell, B. E. Richards, V. Velikov, 
{ J. Phys.: Condens. Matter} {\bf 11,} A75 
(1999).

\bibitem{VMD} 
W. Humphrey, 
A. Dalke, 
K. Schulten, 
{ J. Mol. Graphics} {\bf 14}, 33 
(1996).
 
\end{thebibliography}
\end{document}